\documentclass[10pt,preprint]{aastex}

\newcommand\degree{$^{\circ}$}

\newcommand\simlt{\lower.5ex\hbox{$\; \buildrel < \over \sim \;$}}
\newcommand\simgt{\lower.5ex\hbox{$\; \buildrel > \over \sim \;$}}

\begin{document}
\title{Jet formation in GRBs: A semi-analytic model of MHD flow in Kerr geometry with realistic plasma injection}
\author{Noemie Globus$^{1}$ and Amir Levinson$^{1}$ }
\altaffiltext{1}{School of Physics \& Astronomy, Tel Aviv University, Tel Aviv 69978, Israel}

\begin{abstract}
We construct a semi-analytic model for MHD flows in Kerr geometry, that incorporates energy loading via neutrino annihilation
on magnetic field lines threading the horizon. We compute the double-flow structure for a wide range of energy injection
rates, and identify the different operation regimes. At low injection rates the outflow is powered by the spinning 
black hole via the Blandford-Znajek mechanism, whereas at high injection rates it is driven by the 
pressure of the plasma deposited on magnetic field lines. In the intermediate regime both processes contribute to the outflow formation. 
The parameter that quantifies the load is the ratio of the net power injected below the stagnation radius and the maximum power
that can be extracted magnetically from the black hole. 
\end{abstract} 

\section{Introduction}
An issue of considerable interest in the theory of gamma-ray bursts (GRBs) is the nature of the gamma-ray emitting jet.  
The conventional wisdom has been that the jet is produced by a hyper-accreting black hole that results from a neutron star 
merger in case of short GRBs  (Eichler et al. 1989), or the core-collapse of a massive star in case of long GRBs  \citep{MW99}.
The black hole is likely to be immersed in a strong magnetic field ($B_H\simeq 10^{15}$G) seeded by the progenitor and advected 
inwards during the formation of the central engine. Feedback from a rapidly rotating black hole may dominate the torque experienced by the surrounding 
torus, leading to a state of suspended accretion in long GRBs (van Putten \& Ostriker, 2001; van Putten \& Levinson 2003), provided that mass loading of magnetic field lines anchored to the disk is, somehow,  strongly suppressed (Komissarov \& Barkov 2009; Globus \& Levinson 2013; hereafter GL13).    
Rapid heating of the inner regions of the hyper-accretion disk, or the torus in the suspended accretion state if established, 
leads to prodigious emission of MeV neutrinos (and anti-neutrinos), with  luminosities in the range  $L_\nu=10^{51} - 10^{54}$ erg s$^{-1}$,
depending on accretion rate and specific angular momentum $a$ of the black hole  (e.g., Popham et al. 1999;  Chen \& Beloborodov 2007).   

In the context of the picture outlined above, two competing jet formation mechanisms have been widely discussed in 
the literature; magnetic extraction of the spin down power of a Kerr
black hole, and outflow formation via neutrino  annihilation in the polar region, above the horizon (Paczy\'nski 1990, Levinson \& Eichler 1993, Levinson 2006).  
These two processes are commonly treated under idealized conditions: models of Blandford-Znajek jets usually invoke the force-free limit 
and ignore loading of magnetic field lines (but c.f., Komissarov \& Barkov 2009), whereas models of jets driven by $\nu\bar{\nu}$ 
annihilation (MacFadyen \& Woosley 1999, Fryer \& M\'esz\'aros 2003) are usually constructed within the pure hydrodynamic limit. 
In general, however, both processes might be at work, and it is desirable to characterize the interplay between them.   The approach undertaken 
in this paper is to treat $\nu\bar{\nu}$ annihilation in the magnetosphere as external plasma load.   It has been shown elsewhere (GL13)  that 
injection of relativistically hot plasma on horizon threading field lines always leads to the formation of a double-flow structure in the magnetosphere. 
The plasma inflowing into the black hole carries positive energy that tends to counteract the BZ process. The plasma outflowing to infinity contributes 
to the total asymptotic power. The question addressed in this paper is how the structure of the MHD flow depends on the details of the 
plasma injection process.

In a preliminary investigation (GL13), we considered the effect of the load on the activation of the BZ mechanism,
assuming that the plasma source is confined to an infinitely thin layer, outside which the MHD flow is ideal and adiabatic.  We derived solutions for the inflow section only, and evaluated the critical load above which the BZ process switches off.  We then argued that this critical value
differentiates magnetically extracted from pressure-driven flows when the injected plasma is relativistically hot.  In this paper we 
extend our analysis to more realistic situations, and obtain solutions of the MHD equations for the entire double-flow structure. 
We focus on the conditions anticipated in GRBs, and employ a realistic injection profile computed recently by Zalamea \& Beloborodov  (2011, hereafter ZB11).   We identify the 
different operation regimes, including the intermediate regime where the transition from magnetically extracted to pressure driven flows occurs. We also calculate overloaded solutions for which the BZ process is switch off and compare them with pure hydrodynamic flows derived in a previous study (Levinson \& Globus 2013, hereafter LG13).

\section{Model}

The double-flow structure established in the magnetosphere is illustrated in Figure \ref{f1}: plasma inflow into the black hole and outflow to infinity are ejected from a stagnation radius, $r_{st}$, located between the inner and outer light surfaces.  
The plasma consists of relativistically hot $e^{\pm}$ pairs created via annihilation of MeV neutrinos emitted from the surrounding accretion disk.
The exact location of the stagnation surface depends, quite generally, on the energy injection profile, and is treated as 
an eigenvalue of the MHD equations. 
The double flow possesses six critical surfaces, corresponding to the characteristic phase speeds of the three MHD waves propagating in the medium: two slow magnetosonic, two Alfv\'enic and two fast magnetosonic.
We consider an infinitely conducting, stationary and axisymmetric flow.  
In general, the flow is characterized by a stream function $\Psi(r,\theta)$ that defines the geometry of magnetic flux surfaces, and by the following functionals of $\Psi$: 
the angular velocity of magnetic field lines $\Omega_F(\Psi)$, the  ratio of mass and magnetic fluxes $\eta(\Psi)$, and the energy, angular momentum and entropy  per baryon, denoted by  ${\cal E}(\Psi)$,  ${\cal L}(\Psi)$ and $s(\Psi)$, respectively.
These quantities are given explicitly in Equations (\ref{Enrg})-(\ref{eta}).

The ideal MHD condition implies that $\Omega_F(\Psi)$ is conserved along magnetic flux tubes, as usual. 
All other quantities change along streamlines, owing to plasma injection by the external source, according 
to Equations (\ref{e-flux-derv})-(\ref{eta-derv}). We assume that in the acceleration zone  the plasma 
is relativistically hot with a negligible baryonic content.  
We can therefore adopt the equation of state $w=\rho c^2\bar{h}=4p$.    
To simplify the analysis we invoke a split monopole configuration for the magnetic field lines, described by a stream function of the form 
$\Psi(r,\theta)=\Psi_0(1-\cos\theta)$.  The energy, angular momentum and entropy fluxes, Equation (\ref{MHD-fluxes}), then have only  a radial component:
$\epsilon^r=\rho {\cal E}u^r$, $l^r=\rho {\cal L}u^r$, $s^r=\rho s u^r$. 
Note that ${\cal E}$,  ${\cal L}$ and $s$ diverge in the limit $\rho\rightarrow 0$ (baryon-free flow), 
whereas the corresponding fluxes, $\epsilon^r$, $l^r$ and $s^r$, remain finite and are well defined also in the baryon-free case.
For the relativistic equation of state adopted above the entropy per unit volume is given by $S=\rho s=4p/kT\propto p^{3/4}$, whereby $s^r\propto u^rp^{3/4}$, hence the pressure $p$ can be used as a free variable instead of $S$.  
Since our analysis encompases the force-free limit, we find it convenient to use $\epsilon^r$, $l^r$ and $p$ 
as our free variables.   
With the above simplifications, Equations (\ref{e-flux-derv})-(\ref{s-flux}) reduce to:
\begin{eqnarray}
&&\frac{1}{\sqrt{-g}}\partial_r(\sqrt{-g}\epsilon^r)=-q_t, \label{e1}\\
&&\frac{1}{\sqrt{-g}}\partial_r(\sqrt{-g}\l^r)=q_\varphi,\label{e2}\\
&&\frac{3}{4}\partial_r\ln p=-\partial_r\ln(\Sigma u^r)-\frac{u_\alpha q^\alpha}{4pu^r},\label{e3}
\end{eqnarray}
here $\sqrt{-g}=\Sigma\sin\theta$, and $\Sigma=r^2+a^2\cos^2\theta$.
This set needs to be augmented by an equation of motion for the velocity $u^r$.  Instead of $u^r$ we use the poloidal 
velocity $u_p=\sqrt{u^ru_r}=\sqrt{g_{rr}}u^r$.   Its rate of change along streamlines is derived in appendix \ref{appA} 
and can be written in the form,
\begin{equation}
\partial_r \ln {u_p}=\frac{N_{ad}+N_q}{D},\label{e4}
\end{equation}
where $N_{ad}$, $N_q$ and $D$ are functionals of  $\epsilon^r, l^r, p, u_p, \Omega_F$,  given explicitly by Equations (\ref{D-app})-(\ref{Nq-app}).
Equations (\ref{e1})-(\ref{e4}) form a complete set that governs the structure of the double MHD flow.  
The solutions for the radial profiles of the free variables $\epsilon^r, l^r, p$, and $u_p$,
 depend on the particular choice of the angle $\theta$ that characterizes magnetic flux surfaces.  
 The angular velocity  $\Omega_F(\theta)$ is given as an input.  The energy and angular momentum flow rates 
per solid angle (along a particular flux surface) are defined, respectively, as: 
\begin{equation}
\dot{\cal E}(r,\theta)\equiv\Sigma\epsilon^r,\quad \dot{\cal L}(r,\theta)\equiv \Sigma l^r.\label{angular-power}
\end{equation}

\subsection{\label{source-terms}Source terms for the process $\nu\bar{\nu}\rightarrow e^{+}e^{-}$}

The energy-momentum deposition rate by the reaction $\nu\bar{\nu}\rightarrow e^{+}e^{-}$ was computed in a number of
works, under  different simplifying assumptions (Popham et al. 1999, Chen \& Beloborodov 2007, ZB11).  In what follows we use the recent analysis by ZB11 which includes general relativistic effects.   Following 
ZB11 we denote by $Q^{\alpha}_{\nu\bar{\nu}}$ the local energy-momentum deposition rate measured by a zero-angular-momentum observer (ZAMO).   In general, those rates 
are functions of the Boyer-Lindquist coordinates $r$, $\theta$ and $\varphi$, as can be seen from Figures 2 and 3 in ZB11. 
In terms of the metric components defined in appendix \ref{appA}, $g_{\varphi\varphi}=\varpi^2$, $g_{t\varphi}=-\omega g_{\varphi\varphi}$,
$g_{tt}=-\alpha^2+\omega^2g_{\varphi\varphi}$ and $g_{rr}$, we have the following relations between the ZAMO rates $Q^\alpha_{\nu\bar{\nu}}$ and the source terms $q^\alpha$ measured by a distant observer: $\alpha q^t=Q^t_{\nu\bar{\nu}}$,
$\varpi q^\varphi=Q_{\nu\bar{\nu}}^\varphi+\varpi\omega Q^t_{\nu\bar{\nu}}/\alpha$, $\sqrt{g_{rr}}q^r=Q_{\nu\bar{\nu}}^r$.  From that we obtain
\begin{eqnarray}
-q_t&=&\alpha Q^t_{\nu\bar{\nu}}+\varpi\omega Q_{\nu\bar{\nu}}^\varphi,\label{st-qt}\\
q_\varphi&=&\varpi Q^{\varphi}_{\nu\bar{\nu}}, \label{st-qphi}\\
u_\alpha q^\alpha&=&-\alpha u^t Q_{\nu\bar{\nu}}^{t}+\varpi Q_{\nu\bar{\nu}}^{\varphi}(u^\varphi - \omega u^t)+u_pQ^r_{\nu\bar{\nu}}.\label{st-uq}
\end{eqnarray}
The total power deposited in the magnetosphere can be expressed as,
\begin{equation}
{\dot{E}^{tot}_{\nu\bar{\nu}}} =\int_{r\ge r_H}{\left(\alpha Q^t_{\nu\bar{\nu}}+\varpi\omega Q_{\nu\bar{\nu}}^\varphi    \right)\sqrt{-g}dr d\theta d\varphi}\,.
\label{Edot}
\end{equation}
A fit to the numerical results by ZB11 yields: ${\dot{E}^{tot}_{\nu\bar{\nu}}}\simeq 10^{52} \dot{m}_{acc}^{9/4} \,x_{mso}^{-4.8}$ erg s$^{-1}$ for a  black hole mass $M_{BH}=3M_\odot$, and accretion rates (henceforth measured in units of $M_{\odot}$ s$^{-1}$)  in the range  $0.02<\dot{m}_{acc}<1$, where $x_{mso}$ is the radius of the marginally stable orbit in units of $m=GM_{BH}/c^2$.

Unfortunately, ZB11 do not exhibit results for the azimuthal term $Q^\varphi_{\nu\bar{\nu}}$.  It is also difficult to fit their result for $Q^r_{\nu\bar{\nu}}$.
We shall therefore set $Q^\varphi_{\nu\bar{\nu}}=Q^r_{\nu\bar{\nu}}=0$.   This should not alter much $q_t$ and $u_\alpha q^\alpha$, as 
the first term on the right hand side of Equations (\ref{st-qt}) and (\ref{st-uq}) dominates anyhow.   However, for this choice $q_\varphi=0$, implying 
that the angular momentum flow rate, $\dot{\cal L}(r,\theta)\equiv \Sigma l^r$, 
is conserved, as readily seen from Equation (\ref{e2}).

For our radial flow model it is sufficient to use the angle-averaged energy deposition rate. 
We adopt the form
\begin{equation}
Q^t_{\nu\bar{\nu}}(r)=\dot{Q}_0f(x),
\label{Qdot}
\end{equation}
where $x=r/m$ is a fiducial radius, and $f(x)$ is normalized such that $f(1)=1$. From figures 2 and 3 in ZB11 we 
obtain the approximate profile $f(x)\simeq x^{-b}$, with $b={4.5}$ for a black hole spin parameter $\tilde{a}\equiv a/m=0.95$, and $b=3.5$ 
for $\tilde{a}=0$. The injected power per solid angle, from the horizon to a given radius $x$, is then given by 
\begin{equation}
{\dot{\cal E}_{\nu\bar{\nu}}}(x,\theta) = m^3\dot{Q}_0{\int_{x_H}^{x}{\alpha\Sigma f(x^{\prime}) dx^\prime}}.
\label{Etheta_x}
\end{equation}
The cumulative power distribution in the upper hemisphere ($0\le\theta\le \pi/2$) is
\begin{equation}
\dot{E}_{\nu\bar{\nu}}(x) = 2\pi \int_0^{\pi/2}{\dot{\cal E}_{\nu\bar{\nu}}(x,\theta)\sin\theta d\theta}.
\label{E_x}
\end{equation}
It is related to the total power through ${\dot{E}^{tot}_{\nu\bar{\nu}}}={\dot{E}_{\nu\bar{\nu}}}(x=\infty)$.  The amount 
absorbed by the black hole along a particular field line equals the power per solid angle injected in the 
inflow section (between the horizon and the stagnation radius):
\begin{equation}
{\dot{\cal E}^{in}_{\nu\bar{\nu}}}(\theta) ={\dot{\cal E}_{\nu\bar{\nu}}}(x_{st},\theta)= m^3\dot{Q}_0{ \int_{x_H}^{x_{st}}
{\alpha\Sigma x^{-b} dx}}.
\label{Etheta_in}
\end{equation}
The rest, ${\dot{\cal E}^{out}_{\nu\bar{\nu}}}(\theta)=\dot{\cal E}^{tot}_{\nu\bar{\nu}}(\theta)-\dot{\cal E}^{in}_{\nu\bar{\nu}}(\theta)$,
where $\dot{\cal E}^{tot}_{\nu\bar{\nu}}(\theta)\equiv \dot{\cal E}_{\nu\bar{\nu}}(\infty,\theta)$, emerges at infinity.  
The total power intercepted by the black hole in one hemisphere is 
\begin{equation}
\dot{E}^{in}_{\nu\bar{\nu}}= 2\pi \int_0^{\pi/2}{\dot{\cal E}^{in}_{\nu\bar{\nu}}}(\theta)\sin\theta d\theta =\dot{E}_{\nu\bar{\nu}}(x_{st}).\label{E_in}
\end{equation}
A plot of  $\dot{E}_{\nu\bar{\nu}}(x)$ (Equation (\ref{E_x})), is exhibited in Figure \ref{f2}.   
For our computations we use $b=4.5$ for a spin parameter $\tilde{a}=0.95$ (the solid line in Figure \ref{f2}).

\subsection{The load parameter $\kappa(\theta)$}

Let us denote by $\dot{\cal E}_H(\theta)=\dot{\cal E}(r_H,\theta)$, $\dot{\cal E}_{st}(\theta)=\dot{\cal E}(r_{st},\theta)$,
and $\dot{\cal E}_{\infty}(\theta)=\dot{\cal E}(\infty,\theta)$ the angular distribution of the power at
the horizon, stagnation radius and infinity, respectively, where $\dot{\cal E}(r,\theta)$ is defined in 
Equation (\ref{angular-power}).  Integration of Equation (\ref{e1}) yields 
 \begin{eqnarray}
&&\dot{\cal E}_H(\theta)=\dot{\cal E}_{st}(\theta)-\dot{\cal E}^{in}_{\nu\bar{\nu}}(\theta),\label{EH}\\
&&\dot{\cal E}_{\infty}(\theta)=\dot{\cal E}_{st}(\theta)+\dot{\cal E}^{out}_{\nu\bar{\nu}}(\theta),\label{Einfty}
\end{eqnarray} 
where Equations (\ref{st-qt}), (\ref{Qdot}) and (\ref{Etheta_in}) have been employed.

Now, the specific energy of the injected plasma is positive, hence $\dot{\cal E}^{in}_{\nu\bar{\nu}}(\theta)\ge0$, 
$\dot{\cal E}^{out}_{\nu\bar{\nu}}(\theta)\ge0$, as can be inferred from Equation (\ref{Etheta_x}).  In the situations envisaged here $\dot{\cal E}_{\infty}(\theta)>0$,
but $\dot{\cal E}_{H}(\theta)$ can be negative or positive, depending on the load.  In the force-free limit
$\dot{\cal E}^{tot}_{\nu\bar{\nu}}(\theta)\rightarrow 0$, whereby Equations (\ref{EH}) and (\ref{Einfty}) yield  
$\dot{\cal E}_{\infty}(\theta)=\dot{\cal E}_{H}(\theta)>0$. The extracted power per solid angle is given, in this limit, by (Blandford \& Znajek, 1977): 
\begin{equation}
\dot{\cal E}_H(\theta)=
P_{BZ}(\theta)\equiv \frac{c}{64\pi^3}\alpha_\Omega(1-\alpha_{\Omega})\left(\frac{\tilde{a}}{m}\right)^2\frac{(x_H^2+\tilde{a}^2)\sin^2\theta}{x_H^2(x_H^2+\tilde{a}^2\cos^2\theta)}\Psi_0^2\,,
\label{eflux-ff}
\end{equation}
in terms of the black hole spin $\tilde{a}$, magnetic flux $\Psi_0$, and the dimensionless 
parameter $\alpha_\Omega = \Omega_F/\omega_H$.
In general $\dot{\cal E}_{H}(\theta)<P_{BZ}(\theta)$, as readily seen from Equation (\ref{EH}).

Henceforth, we shall quantify the load on a specific streamline $\theta$ by the parameter
\begin{equation}
\kappa(\theta)\equiv\frac{{\dot{\cal E}^{in}_{\nu\bar{\nu}}}(\theta)}{P_{BZ}(\theta)}.\label{kapp-load}
\end{equation}
At $\kappa(\theta)<<1$ the flow along the streamline $\theta$ is nearly force-free. At $\kappa(\theta)>>1$ the flow 
is nearly hydrodynamic, whereby  $\dot{\cal E}_{st}(\theta)\simeq 0$
and $\dot{\cal E}_H(\theta)\simeq -\dot{\cal E}^{in}_{\nu\bar{\nu}}(\theta) <0$, namely the energy injected below the stagnation radius on
the flux surface $\theta$ is completely absorbed by the black hole.  As shown below, the transition between the two regimes occurs, in general, at $\kappa(\theta)\simeq 1$.

\section{Integration method}

In general, the double flow must pass smoothly through 6 critical surfaces.  Solutions that satisfy this 
requirement can be obtained, in principle, only if the 
trans-field equation is solved simultaneously with Equations (\ref{e-flux-derv})-(\ref{s-flux}), as the exact location of
the critical surfaces is contingent upon the actual shape of the magnetic surfaces.   Such an analysis 
is beyond the scope of this paper.  Fixing the geometry of magnetic field lines renders the system of MHD equations, Eqs. (\ref{e1})-(\ref{e4}),
over constrained.   The reason is that there are 4 regularity conditions (the regularity conditions at the inner and outer Alfv\'en surfaces 
are automatically satisfied), but only 3 adjustable parameters; the location of the stagnation point, $r_{st}$, at which $u_p=0$, 
and the values of $\epsilon^r$ and $p$ at $r_{st}$, henceforth denoted by $\epsilon^r_{st}$, and $p_{st}$, respectively.  
The value of the angular momentum flux
at $r_{st}$ is related to $\epsilon^r_{st}$ through the Bernoulli condition,  Equation (\ref{reg-cond}): $l^r_{st}=\Omega_F^{-1} \epsilon^r_{st}$.  
Since we are interested in determining the energy flux on the horizon, we seek solutions that are  
regular on all 3 inner surfaces, and on the outer slow magnetosonic  surface, but not necessarily on the outer fast magnetosonic surface.  In practice we find 
that any solution that crosses the outer slow magnetosonic surface, is also regular on the outer Alfv\'en surface. 
 
Our strategy is to start with some initial guess for the three adjustable parameters, $r_{st}$, $\epsilon^r_{st}$, and $p_{st}$, whereupon 
Equations (\ref{e1})-(\ref{e4}) are integrated numerically from $r_{st}$ inwards to the horizon, and outwards to the outer Alfv\'en surface, 
along a given field line $\theta$.    The integration 
is repeated many times, where in each run the values $r_{st}$, $\epsilon^r_{st}$,  $p_{st}$ are readjusted until a solution that
crosses the desired critical points smoothly is obtained. In all the examples presented below we neglected the change in linear and angular 
momentum owing to plasma injection, that is, we set $Q_{\nu\bar{\nu}}^\varphi=Q_{\nu\bar{\nu}}^r=0$ in Equations (\ref{st-qt})-(\ref{Edot}).  
Equation (\ref{e2}) then readily yields $\dot{\cal L}=\Sigma l^r=$ const.  
The black hole spin $\tilde{a}$, angular velocity of the streamlines $\Omega_F$, and load parameter $\kappa$ 
are given as input parameters.  In practice, however, $\kappa$ cannot be determined a priori, since $r_{st}$ is unknown.  We therefore use instead of $\kappa$ the 
dimensionless parameter $\tilde{p}_{B}\equiv \Psi_0^2 c/(32\pi^3\dot{Q}_0m^5)$ as an indicator for the load.  Once a solution is obtained and $r_{st}$ is determined, 
$\kappa$ is computed by employing Equations (\ref{Etheta_in}), (\ref{eflux-ff}) and (\ref{kapp-load}). 

\section{Results} 

The family of solutions can be divided into two classes that are distinguished by the sign of the energy flux on the horizon, $\epsilon^r_H$.  
This devision is dictated by the load parameter $\kappa(\theta)$, as discussed further below (Figure \ref{f4}).  We find that in case of underloaded solutions, defined as those for 
which $\kappa(\theta)<<1$, the specific energy is negative in the entire region encompassed by the plasma inflow (below the stagnation radius),
including the horizon, whereby  $\epsilon^r_H>0$. 
In case of overloaded solutions ($\kappa(\theta)>>1$) we find $\epsilon^r_H<0$.
Interestingly,  the energy flux of overloaded solutions changes sign at some radius below the stagnation point, implying that there is still a region 
where the specific energy of the plasma inflow is negative.
Typical examples are shown in Figure \ref{f3}, where the velocity profiles (left panel) and the corresponding energy fluxes (right panel)  of  underloaded ($\kappa=10^{-5}$) and overloaded ($\kappa=20$) equatorial flows are exhibited, for a black hole  spin parameter $\tilde{a}=0.95$, and an energy deposition profile $f(x)=x^{-4.5}$.   In the right panel we also exhibit solutions with $\kappa=0.4$, $1$, $6$, and  $\kappa=\infty$ (a  purely hydrodynamic flow), that are not shown
in the left panel for clarity.    
As seen,  the energy flux of the underloaded solutions is positive everywhere, whereas that of overloaded solutions changes sign below the stagnation radius.   The location $r_0$ at which $\dot{\cal E}(r_0,\theta)=0$ approaches 
$r_{st}$ as $\kappa(\theta)\rightarrow \infty$.
We think that this peculiar behavior stems from the fact that in the regime $\Omega_F<\omega_H$ the Poynting flux measured by a distant observer is always driven by the black hole  (i.e., by frame dragging).  To elucidate this point we employ Equation (\ref{bphi}) to obtain the Poynting flux on the horizon:
\begin{equation}
\left(\frac{F^{r\theta}F_{\theta t}}{4\pi}\right)_H=-\frac{\varpi^2_H\Omega_F(\omega_H-\Omega_F)}{M^2_H+\varpi^2_H(\omega_H-\Omega_F)^2}
(\epsilon^r_H-\omega_H l^r_H).\label{PoynF}
\end{equation}
Now, $\epsilon^r_H-\omega_H l^r_H=\rho_H u_H^r ({\cal E}_H-\omega_H{\cal L}_H)$ is always negative, since the ZAMO energy is always positive, viz.,  ${\cal E}^{ZAMO}={\cal E}-\omega{\cal L}>0$, and $u^r_H<0$.  Thus, for any value of the load parameter, the electromagnetic flux on the horizon is positive if $\omega_H>\Omega_F$.  
Note also that in the force-free limit $M^2_H\rightarrow 0$, and (\ref{PoynF}) reduces to the familiar result, 
$\epsilon^r_H=(F^{r\theta}F_{\theta t})_H/4\pi$, whereas in the pure hydrodynamic case $M^2_H\rightarrow\infty$ and the Poynting flux vanishes, as expected. 
Figure \ref{f4} shows the electric current, $I=\alpha\varpi B_\varphi$, for two overloaded solutions, and it is seen that it never changes sign.  We find that  this is true in general in the regime
$0<\Omega_F<\omega_H$, implying that for any value of $\kappa$ there is a continuous flow of Poynting energy from the horizon outwards, 
against the inflow of injected plasma.   In particular, at the stagnation radius $B_{\varphi }(r_{st})<0$, and from Equation (\ref{Enrg}) we obtain \footnote{This point was not properly understood in GL13. The claim made there, that for overloaded solutions  $B_\varphi$ must vanish at $r_{st}$ is incorrect.  However, the conclusion regarding the activation of the BZ process remains valid, as confirmed in the present analysis.}
\begin{equation}
\epsilon^r_{st}=-\left(\frac{\varpi\Omega_FB_\varphi B_r}{4\pi \sqrt{g_{rr}}}\right)_{st}=\left(\frac{F^{r\theta}F_{\theta t}}{4\pi}\right)_{st}>0.
\end{equation} 
Since for overloaded solutions $\epsilon^r_H<0$, it is evident that the energy flux must vanish at some radius $r<r_{st}$, as seen in the right panel 
of Figure \ref{f3}. 
Our interpretation is that the outward flow of electromagnetic energy driven by the black hole is counteracted by an inward flow of kinetic energy injected on magnetic flux tubes. When the latter exceeds the former the net energy flux becomes negative. For  $\kappa(\theta)<<1$ this never happens, implying energy extraction from the black hole.  For $\kappa(\theta)>>1$ this happens close to the stagnation radius, and since $\epsilon^r_H<0$ we infer that the black hole is being fed by the energy of the overloaded inflow (i.e., the energy injected below the stagnation radius).

Figure \ref{f5}  displays the dependence of the outflow power on the load in the regime where rotational energy extraction 
is switched on, viz., $\epsilon^r_H>0$.   For reference, powers are normalized to the equatorial BZ power, $P_{BZ}(\pi/2)$.  
The horizontal axis gives values of the parameter $\ \dot{\cal E}^{in}_{\nu\bar{\nu}}(\theta)/P_{BZ}(\pi/2)$, which
for the equatorial flow ($\theta=\pi/2$) is just the load parameter $\kappa(\theta)$ defined in Equation (\ref{kapp-load}). For other streamlines, the load parameter
is obtained by multiplying values on the horizontal axis by the factor $P_{BZ}(\pi/2)/P_{BZ}(\theta)$.
The dashed lines delineate the normalized power extracted from the black hole, $\dot{\cal E}_H(\theta)/P_{BZ}(\pi/2)$, and the solid
lines the asymptotic power at infinity, $\dot{\cal E}_{\infty}(\theta)/P_{BZ}(\pi/2)$.    These lines are essentially the locus of solutions obtained from the numerical integration of Equations (\ref{e1})-(\ref{e4}).  Specific cases are indicated by the symbols; 
the circles correspond to solutions for which $\Omega_F=\omega_H/2$ and the triangles to solutions for which $\Omega_F=\omega_H/4$.  As seen,
the effect of the load is highly insensitive to the value of $\Omega_F$, even though the structure of the flow does depend on this parameter (see Figure \ref{f6} below).   This analysis confirms that the transition from underloaded to overloaded flows occurs at $\kappa(\theta)\simeq1$.   We also computed solutions for different black hole spins, and found the same behavior  (see, for example,  Figure 4 in GL13).   One caveat  is the possibility of a nonlinear feedback of the
load on the magnetic flux in the vicinity of the horizon.   Such a feedback may, in principle, change somewhat the activation condition, but not in a 
drastic way. It may be possible to test it using numerical simulations.

As explained above, in our model the angular velocity $\Omega_F$ is given as an input.  In reality it is determined by global conditions. 
For nearly force-free flows numerical simulations indicate that $\Omega_F\simeq \omega_H/2$.  We therefore used this value for the underloaded solutions.  However, when the inertia of the injected plasma becomes important, it is likely to affect $\Omega_F$.  
To study how the properties of the flow depend on this parameter, we sought solutions with different values of $\Omega_F(\theta)$, but the same value of $\kappa(\theta)$.  An example is presented in Figure \ref{f6}, and it is seen that while the velocity profile depends on $\Omega_F$, the power profile is insensitive to the choice of this parameter.    In particular, it does not affect at all the activation condition.  We find this trend is quite general, and therefore conclude that the result exhibited in  Figure \ref{f5}  is robust. 

\section{Conclusion}

We constructed a semi-analytic model for the double-transonic flow established in the magnetosphere of a Kerr black hole under conditions 
anticipated in GRBs, incorporating plasma deposition on magnetic field lines via annihilation of MeV neutrinos emitted by the surrounding
hyper-accretion flow. We examined the effect of energy loading on the properties of the flow, and identified the different operation regimes. 
We find that magnetic extraction of the black hole spin energy ensues, as long as the power deposited below the stagnation radius separating 
the inflow and outflow sections is smaller than the force-free BZ power. The transition from underloaded flows that are powered by the black hole spin energy, to overloaded flows that are powered by the neutrino source  is continuous, as seen in Figure \ref{f5}.

To relate the load parameter derived in Equation (\ref{kapp-load}) to the accretion rate $\dot{m}_{acc}$ (henceforth measured in units of $M_{\odot}$ s$^{-1}$), we employ the scaling relation derived in ZB11.  As mentioned above, their analysis, that exploit an advanced disk model, yields a total energy deposition rate of
\begin{equation}
{\dot{E}^{tot}_{\nu\bar{\nu}}}\simeq 10^{52}\left(M_{BH}/3 M_\odot\right)^{-3/2} \dot{m}_{acc}^{9/4} \,x_{mso}^{-4.8}\quad  {\rm erg \ s^{-1}},
\end{equation}
for accretion rates in the range $0.02<\dot{m}_{acc}< 1$, where $x_{mso}$ is the radius of the marginally stable orbit in units of $m$.    Combining the latter result with the activation condition derived from Figure \ref{f5}, and using the angle averaged energy deposition rate, yields a rough estimate for the accretion rate at which a transition from underloaded to overloaded solutions occurs:
\begin{equation}
\dot{m}_{c}\simeq 1\left(\frac{M_{BH}}{3M_\odot}\right)^{-2/9}\left(\frac{\Psi_0}{10^{28}{\rm \,G\ cm^2}}\right)^{8/9}f(\tilde{a}).\label{f(a)}
\end{equation}
Here $\Psi_0$ is the magnetic flux accumulated in the vicinity of the horizon,  and the function $f(\tilde{a})$ is displayed in Figure \ref{f7}.
According to this relation, when $\dot{m}_{acc}<\dot{m}_c$ 
the outflow is powered by the BZ process, whereas for $\dot{m}_{acc}>\dot{m}_c$ it  is driven by the neutrino source.  In reality, the magnetic flux $\Psi_0$ should also depend on the accretion rate, however, the sensitivity of this relation to the
assumptions underlying the specific disk model adopted for its calculation renders it highly uncertain.  Furtheremore, the 
presence of sufficiently strong magnetic field in the inner disk regions may affect the neutrino luminosity.  
For illustration, we use the disk model of Chen \& Beloborodov (2007) to estimate the magnetic flux.  Unfortunately, it is 
difficult to derive scaling relations from the results presented in this paper, but from Figures 1 and 2 there we
obtained $\Psi_0\sim 2\times10^{28}\sqrt{\xi_B}$ G cm$^2$ 
for a black hole mass $M_{BH}=3M_\odot$, angular momentum $\tilde{a}=0.95$, viscosity parameter $\alpha_{vis}=0.1$, and accretion rate 
$\dot{m}_{acc}=0.2$, assuming that the magnetic pressure in the inner regions of the disk 
is a fraction $\xi_B$ of the total pressure.  For this choice we infer that with $\xi_B$ on the order of a few percents, as naively 
expected, the transition from underloaded to overloaded flows may occur at accretion rates
$\dot{m}_{acc}>0.1$ or so.   We emphasize that this estimate is highly uncertain.  

The above results may also have some implications for the jet structure.  To be concrete, the dependence of the activation condition on the inclination angle $\theta$ of magnetic surfaces (see  Figure  {\ref{f5}), and the
approximate uniformity of the angular distribution of the energy deposition rate indicated in Figures 2 and 3 of ZB11,
suggest that for accretion rates $\dot{m}_{acc}\simlt\dot{m}_c$, and unless the magnetic flux near horizon is extremely high, the outflow produced in the polar region may consist of an inner core inside which the power is dominated by the thermal energy of the hot plasma, and outside which it is dominated by the Poynting flux driven by frame dragging. 

This research was supported by a grant from the Israel Science Foundation no. 1277/13

\appendix
\section{\label{appA}Derivation of the flow equations}

The stress-energy tensor of a magnetized fluid takes the form,
\begin{equation}
T^{\alpha\beta} = \bar{h}\rho c^2u^{\alpha}u^{\beta} + pg^{\alpha\beta} + \frac{1}{4\pi}\left( F^{\alpha\gamma}F^{\beta}_{\gamma}-\frac{1}{4}g^{\alpha\beta}F^2 \right),
\label{T_M}
\end{equation}
here $u^{\alpha}$ is the four-velocity measured in units of c, $\bar{h}=(\rho c^2+ e_{int} +p) /{\rho c^2}$ the dimensionless specific enthalpy, $\rho$ the baryonic rest-mass density, $p$ the pressure, and $g_{\mu\nu}$ the coefficients of the metric tensor of the Kerr spacetime. In the following we use geometrical units ($c=G=1$), unless otherwise stated, and express the Kerr metric in the regular Boyer-Lindquist coordinates,
$ds^{2} \equiv g_{\mu\nu}dx^{\mu}dx^{\nu}$
with the non-zero metric coefficients given by: $g_{rr}={{\Sigma}/{\Delta}}$, $g_{\theta\theta}={\Sigma}$, $g_{\varphi\varphi}\equiv\varpi^2={A}\sin^2\theta/\Sigma$, $g_{tt}=-\alpha^2+\omega^2g_{\phi\phi}$, $g_{t\phi}=-\omega g_{\phi\phi}$,  in terms of $\Delta=r^{2}+a^{2}-2mr$, $\Sigma=r^{2}+a^{2}\cos^{2}\theta$, $A=(r^{2}+a^{2})^{2}-a^{2}\Delta \sin^{2}\theta$, $\alpha=\sqrt{\Sigma\Delta/A}$, and $\omega=2mra/A$. The parameters $m$ and $a$ are the mass and specific angular momentum per unit mass of the hole, $\alpha$ is the time lapse and $\omega$ the frame dragging potential between a zero-angular-momentum observer (ZAMO) and an observer at infinity.  The angular velocity of the black hole is defined as the value of $\omega$ on the horizon, viz.,  $\omega_H\equiv \omega(r=r_H)=a/(2mr_H)$, here $r_H=m+\sqrt{m^2-a^2}$ is the radius of the horizon, obtained from the condition $\Delta_H=0$.

The dynamics of the flow is governed by the energy-momentum equations:
\begin{equation}
\frac{1}{\sqrt{-g}}(\sqrt{-g}T^{\alpha\beta})_{,\alpha}+
\Gamma^{\beta}_{\ \mu\nu}T^{\mu\nu} = q^\beta, \label{Ttot=q}
\end{equation}
mass conservation:
\begin{equation}
\frac{1}{\sqrt{-g}}\partial_\alpha(\sqrt{-g}\rho u^\alpha)=q_n,\label{continuity-eq}
\end{equation}
and Maxwell's equations:
\begin{eqnarray}
F^{\beta\alpha}_{;\alpha}=\frac{1}{\sqrt{-g}}(\sqrt{-g} F^{\beta\alpha})_{,\alpha}=4\pi j^\beta,\\
F_{\alpha\beta,\gamma}+F_{\beta\gamma, \alpha}+ F_{\gamma\alpha,\beta}=0,\label{Maxwell-homg}
\end{eqnarray}
subject to the ideal MHD condition $F^{\mu\nu}u_\nu=0$.  Here, $q^{\beta}$ denotes the source terms associated with energy-momentum transfer by an external agent, $q_n$ is a particle source, 
and $\Gamma^\beta_{\mu\nu}$ denotes the affine connection.    

The energy, angular momentum and entropy fluxes, can be expressed explicitly as 
\begin{equation}
\epsilon^a\equiv -T^a_t=\rho u^a{\cal E},\quad l^a\equiv T^a_\varphi=\rho u^a{\cal L},\quad s^a=(\rho/m_N) u^a s,\label{MHD-fluxes}
\end{equation}
in terms of the energy per baryon,
\begin{eqnarray}
{\cal E}= -\bar{h}u_t - \frac{\alpha\varpi\Omega_F}{4\pi\eta}B_{\varphi},\label{Enrg}
\end{eqnarray}
angular momentum per baryon,
\begin{eqnarray}
{\cal L}= \bar{h}u_\varphi -  \frac{\alpha\varpi B_{\varphi}}{4\pi\eta },\label{Lmom}
\end{eqnarray}
and the entropy per baryon $s$, where
\begin{equation}
\Omega_F= v^\varphi-\frac{v_pB_\varphi}{\varpi B_p}\label{Omega}
\end{equation}
is the angular velocity of magnetic field lines,
\begin{equation}
\eta=\frac{\rho u_p}{B_p}\label{eta}
\end{equation}
is the  ratio of mass and magnetic fluxes, and the index $a$ runs over $r$ and $\theta$. 
In the above equations $u_p=\pm(u_r u^r+u_\theta u^\theta)^{1/2}$ is the poloidal velocity, where the plus sign applies to outflow lines and the minus sign to inflow lines, $v_p=u_p/\gamma$, with $\gamma=u^t\alpha$ being the 
Lorentz factor measured by a ZAMO, $v^\varphi=u^\varphi/u^t$,
 $B_p=(B_r^2+B_\theta^2)^{1/2}/\alpha$ is the redshifted poloidal magnetic field, and
$B_r=F_{\theta\varphi}/\sqrt{A}\sin\theta$,  $B_\theta=\sqrt{\Delta}F_{\varphi r}/\sqrt{A}\sin\theta$ and $B_\varphi=\sqrt{\Delta}F_{r\theta}/\Sigma$ the magnetic field components measured by a ZAMO (see van Putten and Levinson 2012 for details).
Note that with our sign convention the value of $\eta$ is 
defined to be positive on outflow lines and negative on inflow lines.

For a stationary and axisymmetric ideal MHD flow, Equations (\ref{Ttot=q})-(\ref{Maxwell-homg}) can be
reduced to (GL13)
\begin{eqnarray}
&&\frac{1}{\sqrt{-g}}\partial_a(\sqrt{-g}\epsilon^a)=-q_t, \label{e-flux-derv}\\
&&\frac{1}{\sqrt{-g}}\partial_a(\sqrt{-g}\l^a)=q_\varphi,\label{L-flux-derv}\\
&&\frac{kT}{\sqrt{-g}}\partial_a(\sqrt{-g}s^a)=-u_a q^\alpha,\label{s-flux}\\
&&u^a\partial_a\eta=\frac{u_pq_n}{B_p},\label{eta-derv}\\
&&u^a\partial_a\Omega(\Psi)=0.
\end{eqnarray}
It can be readily seen that for $q_n=q^\mu=0$, the quantities $\Omega(\Psi)$, ${\cal E} (\Psi)$,
${\cal L} (\Psi)$, $\eta(\Psi)$ and $s(\Psi)$ are conserved on magnetic flux surfaces $\Psi(r,\theta)=$const, as
is well known (e.g., Camenzind 1986).

From (\ref{Enrg})-(\ref{eta}) we obtain the expressions: 
\begin{eqnarray}
&& B_{\varphi}=-\frac{4\pi\eta{\cal E}}{\alpha\varpi} \;\frac{\alpha_{}^{2}\tilde{L}-\varpi^{2}(\Omega_F-\omega)(1-\tilde{L}\omega)}{k_0-M^{2}}\label{bphi}\,\\
&& u^t = \frac{{\cal E}}{\bar{h}}\,\frac{\alpha^2(1-\Omega_F{\tilde{L}})-
M^2(1-\omega{\tilde{L}})}{\alpha^2\left(k_0-M^2\right)},\\
&& u^\varphi=\frac{{\cal E}}{\bar{h}}\frac{\alpha^2 \Omega_F({1-\Omega_F\tilde{L})
-M^2\omega(1-\omega\tilde{L})-{M^2\tilde{L}\alpha^2}{\varpi^{-2}}}}{\alpha^2\left(k_0-M^2\right)}\,,\label{uphi}
\end{eqnarray}
here $\tilde{L}={\cal L}/{\cal E}$,  $M$ as the poloidal Alfv\'enic Mach number, defined through $M^2\equiv{4\pi \bar{h} {\eta}^2 c^2}/{\rho}={u_p^2}/{u_A^2}$, and $u_A^2=B_p^2/(4\pi \bar{h} \rho c^2)$.  Combining the latter relations with the normalization condition $u^\alpha u_\alpha=-1$
yields the Bernoulli equation (Camenzind, 1986; Takahashi et al. 1990):
\begin{equation}
{u}_p^2+1=\left(\frac{{\cal E}}{\bar{h}}\right)^2\frac{k_0 k_2 - 2 k_2 M^2 -k_4 M^4}{(k_0-M^2)^2}\label{mot}\,,
\end{equation}
where 
\begin{eqnarray}
k_0 &=& \alpha^2 - \varpi^2\left(\Omega_F-\omega\right)^2,\\
k_2 &=& (1-{\tilde L}\Omega)^2,\\
k_4 &=& \frac{{\tilde L}^2}{\varpi^2}-\frac{(1-{\tilde L}\omega)^2}{\alpha^2}\label{k_4}\,.
\end{eqnarray}
 In terms of the free variables, $\epsilon^r$, $l^r$, $p$, used in our integration we have: $\eta{\cal E}=\sqrt{\Sigma}\epsilon^r/(\sqrt{\Delta}B_p)$, and 
${\cal E}/\bar{h}=\epsilon^r/(4pu^r)$, $\tilde{L}=l^r/\epsilon^r$.

The equation of motion (\ref{e4})  is obtained upon differentiating Equation (\ref{mot}) along a given streamline,
using Equations (\ref{e-flux-derv})-(\ref{s-flux}) with the source terms $q_t=-\alpha \dot{Q}_0 f(x)$, $q_\phi=0$ and $u_\alpha q^\alpha=u^t q_t$, which are derived in section \ref{source-terms}, and noting that in the split monopole geometry, the redshifted poloidal field reduces to $B_p = \Psi_0/(2\pi \sqrt{\Sigma\Delta})$, the poloidal velocity is $u_p=\sqrt{\Sigma/\Delta}u^r$, and the convective derivative reduces to $u^\alpha\partial_\alpha = u^r\partial_r=\sqrt{\Delta/\Sigma}u_p\partial_r$.   This yields the following expressions  for the functionals $D$, $N_{ad}$ and $N_q$ in Equation (\ref{e4}):
\begin{eqnarray}
D=\left(k_0-M^2\right)^2\left[\left(u_p^2-c_s^2\right)\left(k_0-M^2\right)+ \left(\frac{{\cal E}}{\bar{h}}\right)^2 M^4\frac{\left(k_0 k_4+k_2\right)}{\left(k_0-M^2\right)^2} \right]\,,\label{D-app}\\
N_{ad}=\left[-\left(1+{u}_p^2\right)\left(k_0-M^2\right)^3c_s^2+\left(\frac{{\cal E}}{\bar{h}}\right)^2  M^4\left(k_0 k_4 + k_2\right)\right]\partial_x\ln {B}_p \\
\nonumber - \frac{3}{16}   \left(\frac{{\cal E}}{\bar{h}}\right)^2 \left[M^4\left(k_0-M^2\right)\partial_x k_{ad}+\left(k_0 k_2 - 3 k_2 M^2 - 2 k_4 M^4\right)\partial_xk_0\right]\,,\label{Nad-app}\\
N_q = -\frac{3\, q_t}{2\, \epsilon^r}\left(\frac{{\cal E}}{\bar{h}}\right)^2(k_0-M^2)\left[\left(k_0-2M^2\right)(1-\Omega_F\tilde{L})+\frac{M^4}{\alpha^2}(1-\omega{\tilde{L}})\right]\nonumber\\
-\frac{2\, q_t}{\epsilon^r}\left(\frac{{\cal E}}{\bar{h}}\right)^4\left[-k_4M^6-k_2\left(k_0^2-3k_0M^2+3M^4\right)\right]\frac{\alpha^2(1-\Omega_F{\tilde{L}})-M^2(1-\omega{\tilde{L}})}{\alpha^2\,(k_0-M^2)},\label{Nq-app}
\end{eqnarray}
where the derivatives are defined by
\begin{eqnarray}
\partial_xk_0 &=& \partial_x(\alpha^2) - \left(\Omega_F-\omega\right)^2 \partial_x(\varpi^2)+2\varpi^2\left(\Omega_F-\omega\right)\partial_x\omega,\\
\partial_x k_{ad}&=&\frac{2}{\alpha^2}(1-{\tilde L}\omega)^2\partial_x\ln\alpha+\frac{2{\tilde L}\omega}{\alpha^2}(1-{\tilde L}\omega)\partial_x\ln\omega-\frac{2{\tilde L}^2}{\varpi^2}  \partial_x\ln\varpi\,.
\end{eqnarray}

\subsection{The stagnation point}\label{app_reg}

At the stagnation point $x=x_{st}$, where $u_p=0$, Equations (\ref{e1})-(\ref{e4}) with $q_\varphi=0$ yield:
\begin{eqnarray}
\partial_x {u_p}_{|_{x=x_{st}}}&=&-\frac{(\Sigma q_t)_{st}}{4 {p}_{st}\alpha_{st}(k_{0st}\,A_{st})^{1/2}},\\
\partial_x(\Sigma\epsilon^r)_{|_{x=x_{st}}}&=&-(\Sigma q_t)_{st}\,,\\ 
\frac{3}{4}\partial_x\ln(p)_{|_{x=x_{st}}}&=&-\frac{x_{st}-1}{\Delta_{st}}-\frac{x_{st}}{\Sigma_{st}}-\frac{\Sigma_{st}^2 f(x_{st})\epsilon^r_{st}(1-\omega_{st}/\Omega_F)}{8p_{st}\alpha_{st}k_{0st}\tilde{p}_{B}},
\label{stagnation}
\end{eqnarray}
where $-(\Sigma q_t)_{st}=\alpha_{st}\Sigma_{st}\dot{Q}_0f(x_{st})$, and the parameter $\tilde{p}_{B}\equiv \Psi_0^2 c/(32\pi^3\dot{Q}_0m^5)$ is an indicator for the load.

The Bernoulli condition can be rewritten
\begin{equation}
(1-{\tilde L}_{st}\Omega_F) \frac{\epsilon^r_{st}}{(4pu^r)_{st}}=\sqrt{k_{0st}}\,,
\label{reg-cond}
\end{equation}
implying ${\tilde L}_{st}\Omega_F =1$, since as we have shown,  there is always extraction of angular momentum from the black hole so that $\epsilon^r_{st}>0$.

\subsection{The Alfv\'en surfaces}

The location  of the Alfv\'en surfaces is defined by the roots of the denominator in Equations (\ref{bphi})-(\ref{uphi}), 
$M_A^2=k_{0A}= \alpha_A^2-\varpi_A^2(\Omega_F-\omega_A)^2$, here the subscript $A$ denotes values on this surface.
The latter equation has two roots, corresponding to the inner and outer Alfv\'en surfaces.  
The requirement that $B_\varphi$, $u^t$ and $u^\varphi$ in Equations (\ref{bphi})-(\ref{uphi}) are continuous there imposes a condition on the ratio
of the ZAMO energy, ${\cal E}^{ZAMO}={\cal E}-\omega{\cal L}$, and the energy ${\cal E}$ of an observer at infinity: 
\begin{equation}
\left(\frac{{\cal E}}{{\cal E}^{ZAMO}}\right)_A=1-\alpha_A^{-2}\varpi_A^2\omega_A(\omega_A-\Omega_F). \label{EZAMO/E}
\end{equation}
In the regime where the BZ process is activated, $\epsilon_H^r=\rho u^r{\cal E}>0$,  and we must have ${\cal E}<0$ anywhere below the stagnation radius
where $u^r<0$, and in particular at the inner Alfv\'en point ($IA$).  Since the ZAMO energy is always positive, the latter condition, combined with Equation (\ref{EZAMO/E}), readily implies $\omega_{IA}>\Omega_F$, and defines the range of Alfv\'en radii that are permitted for energy extraction:
 \begin{equation}
\frac{\alpha_{IA}}{\sqrt{\omega_{IA}(\omega_{IA}-\Omega_F)}}<\varpi_{IA}.
\end{equation}
The shaded area in Figure \ref{f8} marks this range for solutions
with $\Omega_F=\omega_H/2$.   In the outflow section all energies are positive, yielding $\omega_{OA}<\Omega_F$, and 
 \begin{equation}
\frac{\alpha_{OA}}{\sqrt{\omega_{OA}(\omega_{OA}-\Omega_F)}}>\varpi_{OA},
\end{equation}
at the outer Alfv\'en point ($OA$).

\section{Pure hydrodynamic flows}
The equations governing a purely hydrodynamic flow can be obtained formally from the MHD equations derived above upon taking the limit  $M^2\rightarrow\infty$:
\begin{eqnarray}
(u_p^2-c_s^2)\partial_x\ln u_p=N_{ad}+N_q,\label{eq-motion1}\\
N_{ad}=(1+u_p^2)c_s^2\left(\frac{x}{\Sigma}+\frac{x-1}{\Delta}\right)-\left(\frac{{\cal E}}{\bar{h}}\right)^2\frac{3\partial_x k_{ad}}{4}\,,\label{eq-motion-2}\\
N_q =-\frac{3\,q_t}{2\alpha^2\epsilon^r}(1-{\tilde L}\omega)\left(\frac{{\cal E}}{\bar{h}}\right)^2\left[1+\left(\frac{{\cal E}}{\bar{h}}\right)^2 \frac{4\,k_4}{3}\right]\,.
\label{eq-motion-3}
\end{eqnarray}

\begin{equation}
\frac{3}{4}\partial_x\ln \tilde{p}=-\partial_x\ln u_p-\frac{x}{\Sigma}-\frac{x-1}{\Delta}-\frac{q_t(1-{\tilde L}\omega)}{\alpha^2\epsilon^r}\left(\frac{{\cal E}}{\bar{h}}\right)^2,
\label{eq-pressure}
\end{equation}

\begin{equation}
\partial_x(\Sigma\epsilon^r)=-\Sigma q_t\,,
\label{eq-energyflux}
\end{equation}
where now ${\tilde L}$ is a free parameter that describes the family of solutions, $\tilde{p}=p/\dot{Q}_0 t_d$ is the normalized pressure, with $t_d=GM_{BH}/c^3$.  The Bernoulli condition (\ref{reg-cond}) implies, in this limit, $\epsilon^r_{st}=0$, as expected in the absence of magnetic fields.    It can be readily shown that when $M^2\rightarrow\infty$ the slow-magnetosonic and Alfven  speeds approach zero,
whereas the fast-magnetosonic speed approaches the sound speed, $c_s=1/\sqrt{2}$.  Consequently, the above system of equations has critical points at $u_p=\pm c_s$, as can be directly verified.

The regularity conditions at the sonic points, obtained from Equations (\ref{eq-motion-2}) and (\ref{eq-motion-3}), read:

\begin{equation}
2\left(\frac{x_{c1}}{\Sigma_{c1}}+\frac{x_{c1}-1}{\Delta_{c1}}\right)+3\frac{(\partial_x k_{ad})_{c1}}{k_{4,c1}}=-\sqrt{\frac{3}{-k_{4,c1}}}\frac{\sqrt{\Sigma_{c1}} f(x_{c1})(1-{\tilde L}_{c1}\omega_{c1})}{{\tilde p_{c1}}\sqrt{\Delta_{c1}}\alpha_{c1}},\label{x_c1}
\end{equation}

\begin{equation}
2\left(\frac{x_{c2}}{\Sigma_{c2}}+\frac{x_{c2}-1}{\Delta_{c2}}\right)+3\frac{(\partial_x k_{ad})_{c2}}{k_{4,c2}}=+\sqrt{\frac{3}{-k_{4,c2}}}\frac{\sqrt{\Sigma_{c2}} f(x_{c2})(1-{\tilde L}_{c2}\omega_{c2})}{{\tilde p_{c2}}\sqrt{\Delta_{c2}}\alpha_{c2}},\label{x_c2}
\end{equation}
denoting the sonic point of the inflow (outflow) by $x_{c1} (x_{c2})$, and
noting that $u_p= -1/\sqrt{2}$ at $x_{c1}$, and  $u_p= 1/\sqrt{2}$ at $x_{c2}$.  As seen, the existence of two sonic points, that is, $x_{c1}\ne x_{c2}$, is a consequence of energy injection. When $f(x)=0$ the solutions of (\ref{x_c1}) and (\ref{x_c2}) merge, and the system has only one critical point, for either an inflow or an outflow, depending on the boundary conditions.

At the stagnation point, $x=x_{st}$, the above equations yield 
\begin{eqnarray}
\partial_x {u_p}_{|_{x=x_{st}}}&=&\frac{f(x_{st})\sqrt{\Sigma_{st}}(1-{\tilde L}_{st}\omega_{st})}{4 \tilde{p}_{st}\sqrt{\Delta_{st}}\alpha_{st}\sqrt{-k_{4,st}}},\\
\partial_x(\Sigma\epsilon^r)_{|_{x=x_{st}}}&=&-(\Sigma q_t)_{st}\,,\\ 
\partial_x\ln(\tilde{p})_{|_{x=x_{st}}}&=&\frac{2}{k_{4,st}}(\partial_x k_{ad})_{st},
\label{stagnation}
\end{eqnarray}
where $\tilde{p}_{st}=\tilde{p}(x_{st})$ is the normalized stagnation pressure.  Thus, for a given choice of $f(x)$ the solution is fully determined once $x_{st}$ and $\tilde{p}_{st}$ are known, since $\epsilon^r_{st}=0$.

\newpage

\begin{figure}[ht]
\centering
\includegraphics[width=9.5cm]{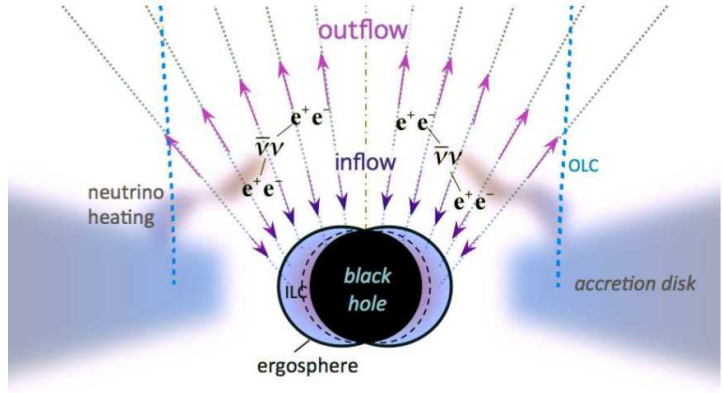}
\caption{\label{f1} Illustration of the double-transonic flow model. }
 \end{figure}

\begin{figure}[ht]
\centering
\includegraphics[width=11cm]{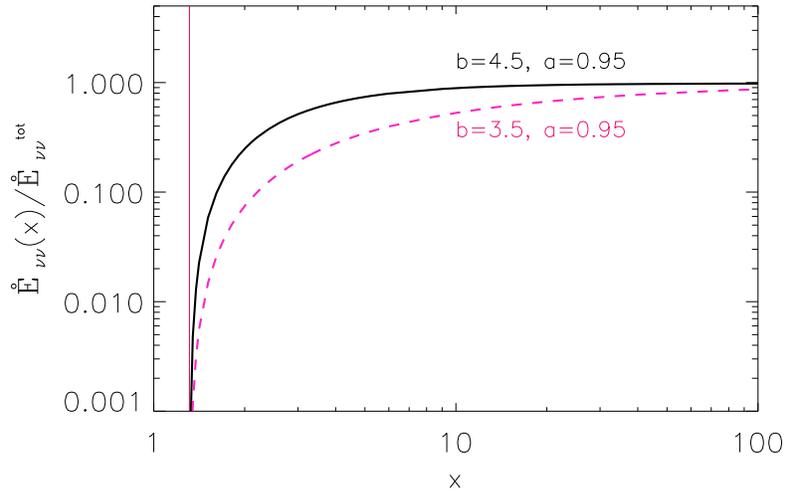}
\caption{\label{f2} Net power deposited on magnetic field lines via neutrino annihilation as a function of 
radius $x$ (Equation (\ref{E_x})), for two different injection profiles, $f(x) = x^{-4.5}$ and $f(x) = x^{-3.5}$ ($\tilde{a}=0.95$).} 
 \end{figure}

\begin{figure}[ht]
\centering
\includegraphics[width=16cm]{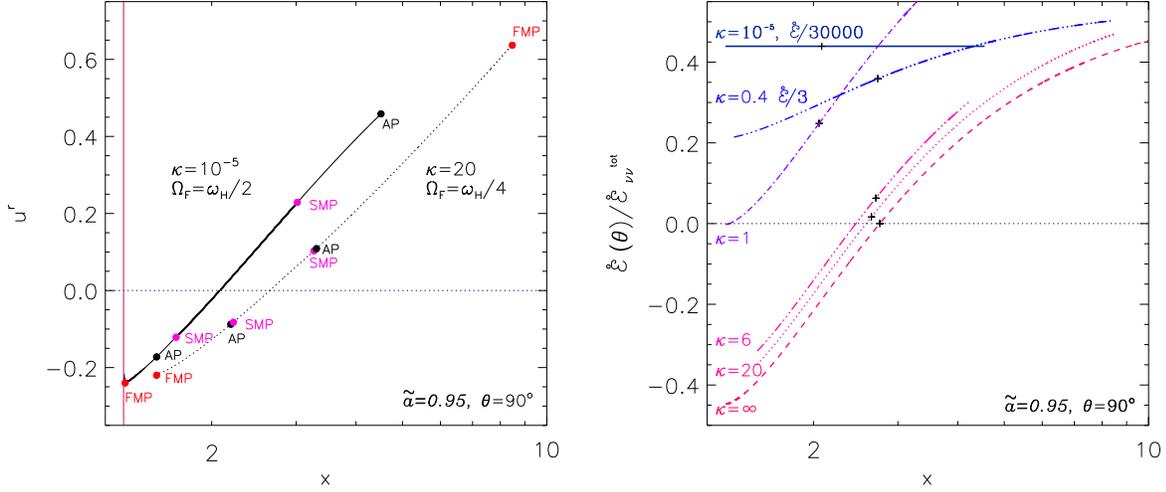}
\caption{\label{f3} Left panel: velocity profiles of underloaded ($\kappa=10^{-5}$) and overloaded ($\kappa=20$) 
solutions, with $f(x) = x^{-4.5}$, $\tilde{a}=0.95$, and $\theta=90$\degree.  The region above (below) the 
horizontal dotted line $u^r=0$, corresponds to the outflow (inflow) sections.  The inner and outer 
slow magnetosonic points (SMP), Alfv\'en points (AP), and fast magnetosonic points (FMP) are indicated.  The vertical red
line delineates the horizon.  Right panel: profiles of the outflow power per solid angle for different values of 
the load parameter $\kappa$.  The $\kappa=10^{-5}$ and $\kappa=0.4$ curves are rescaled for convenience.  The cross symbole on each curve marks the location of the stagnation radius.  In the region of the inflow where the energy flux is positive  (between the point of zero flux and  the stagnation point) the specific energy is negative.}
\end{figure}

\begin{figure}[ht]
\centering
\includegraphics[width=11cm]{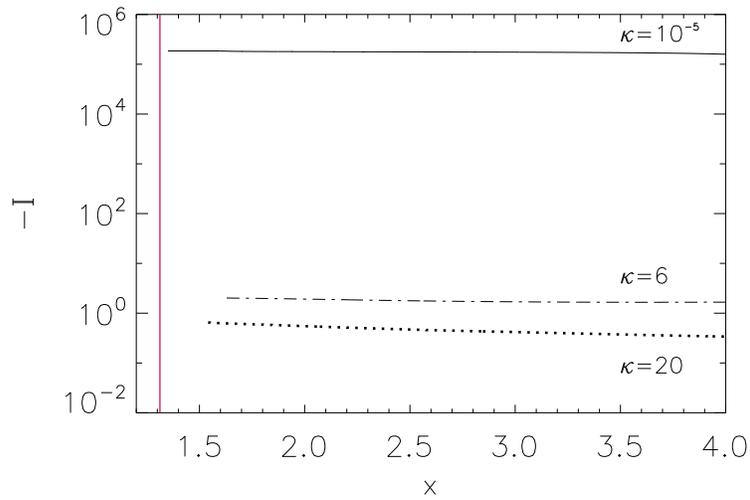}
\caption{\label{f4} Electric current distribution, $I(x)= B_\varphi\varpi\alpha$, for equatorial flow solutions.}
 \end{figure}

\begin{figure}[ht]
\centering
\includegraphics[width=10cm]{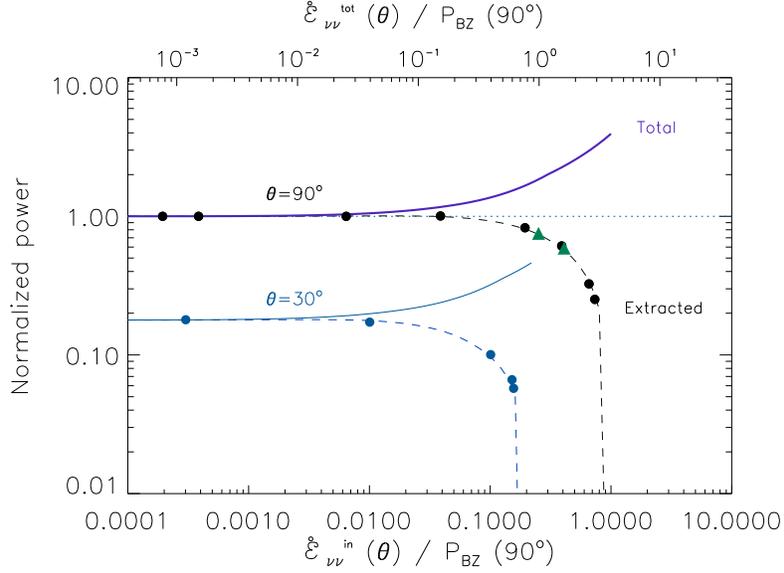}
\caption{\label{f5} Dependence of the outflow power on the load for two different streamlines, 
$\theta=90^\circ$, and $\theta=30^\circ$.  For reference, powers are normalized by the equatorial BZ power, $P_{BZ}(\pi/2)$, given
in Equation (\ref{eflux-ff}).
The dashed line in each case gives the normalized power per solid angle
on the horizon, $\dot{\cal E}_H(\theta)/P_{BZ}(\pi/2)$, and the solid line the total power per solid angle, 
$\dot{\cal E}_\infty(\theta)/P_{BZ}(\pi/2)$.  Specific cases are indicated by the symbols, with the circles 
corresponding to solutions for which $\Omega_F=\omega_H/2$, and the triangles to solutions with $\Omega_F=\omega_H/4$.}
\end{figure}

\begin{figure}[ht]
\centering
\includegraphics[width=16cm]{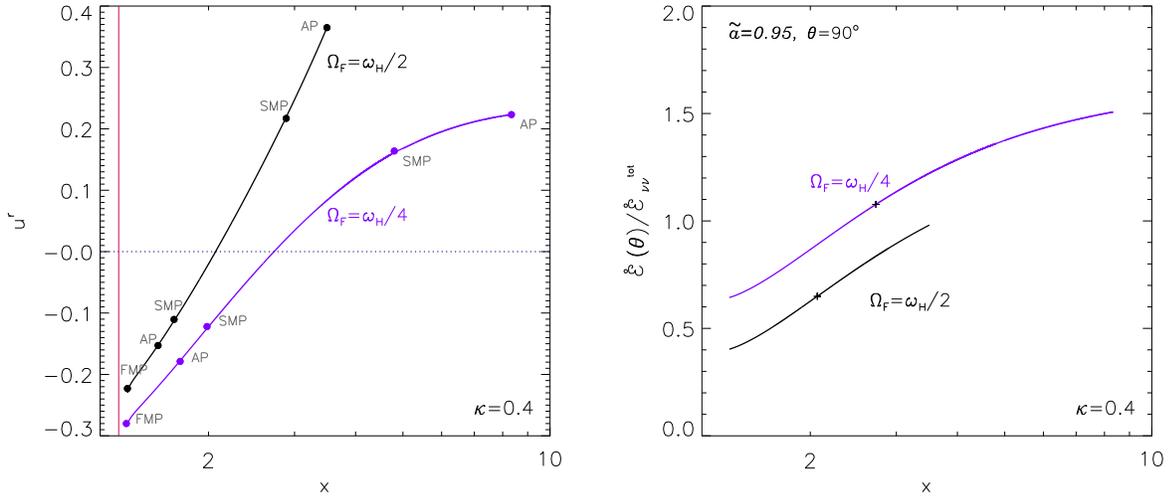}
\caption{\label{f6} A comparison between two equatorial flow solutions with the same load parameter, $\kappa=0.4$,
and different angular velocities $\Omega_F$, as indicated.  The left panel displays the velocity profiles
and the right panel the corresponding power profiles.}
\end{figure}

\begin{figure}[ht]
\centering
\includegraphics[width=11cm]{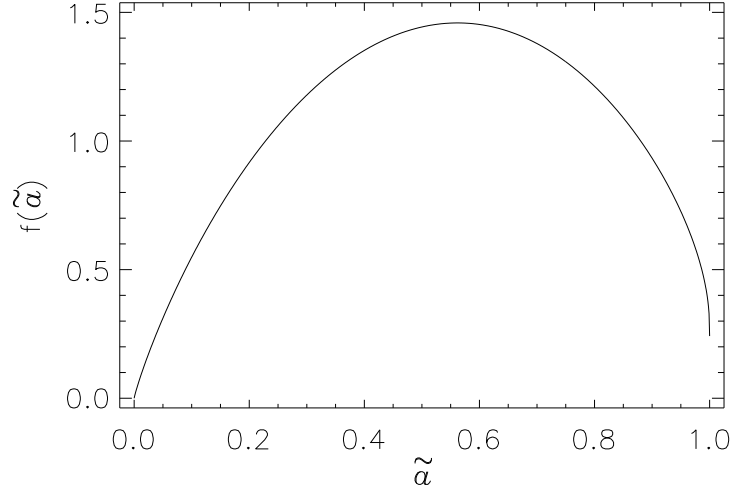}
\caption{\label{f7} A plot of the function $f(\tilde{a})$ defined in Equation (\ref{f(a)}).}
\end{figure}

\begin{figure}[ht]
\centering
\includegraphics[width=11cm]{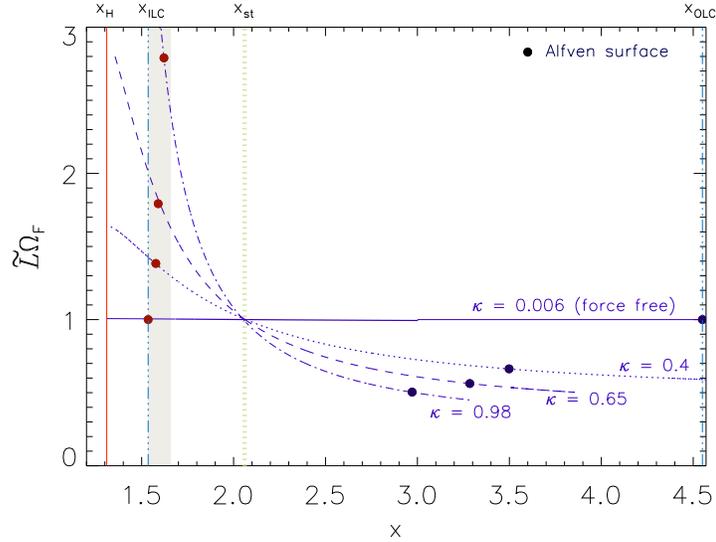}
\caption{\label{f8}Effect of the loading efficiency on the position of the Alfv\'en surfaces. The profile of $\tilde{L}\Omega_F$ is shown for 4 solutions corresponding to different values of $\kappa$, and $\tilde{a}=0.95$,  $\Omega_F=\omega_H/2$. The range of Alfv\'en radii that allows rotational energy extraction is delineated by the shaded area. In the force-free case ($\kappa\rightarrow0$, $\dot{\cal E}_H=P_{BZ}$) the Alfv\'en points coincide with the light surfaces.  This result is in accord with that derived by Takahashi et al. (1990) for an adiabatic flow.} 
\end{figure}

\end{document}